\documentclass[letterpaper, 10 pt, conference]{ieeeconf}
\IEEEoverridecommandlockouts
\usepackage{amsmath,amssymb,amsfonts}
\usepackage{blindtext, graphicx}
\usepackage{cite}
\PassOptionsToPackage{greek,english}{babel}
\usepackage{algorithm}
\usepackage{textcomp}
\usepackage{xcolor}
\usepackage{booktabs} 
\usepackage{bbding}
\usepackage{soul,color}
\usepackage{comment} 
\usepackage{kantlipsum}
\usepackage{multicol}
\usepackage{microtype}
\usepackage{array}
\usepackage{tabulary}
\usepackage{chngcntr}
\usepackage{multirow}
\usepackage[T1]{fontenc}
\usepackage{amssymb}

\usepackage{tablefootnote}
\usepackage{threeparttable}
\usepackage[noend]{algpseudocode}

\usepackage{cleveref}

\let\Setlength\setlength 
\usepackage{calc}
\newlength{\arrayrulewidthOriginal}

\DeclareMathAlphabet\mathzapf{T1}{pzc}{mb}{it}

\newcolumntype{K}[1]{>{\centering\arraybackslash}p{#1}}
\usepackage{float}
\usepackage{bm}

\newcolumntype{M}[1]{>{\centering\arraybackslash}m{#1}}
\Setlength{\intextsep}{5pt}
\renewcommand{\baselinestretch}{1} 

\IEEEoverridecommandlockouts                              

\algdef{SE}[SUBALG]{Indent}{EndIndent}{}{\algorithmicend\ }%
\algtext*{Indent}
\algtext*{EndIndent}

\newcommand{\systemname}{ReBeatICG}
\begin{document}
\title {\systemname{}: Real-time Low-Complexity Beat-to-beat\\Impedance Cardiogram Delineation Algorithm\\
\thanks{This work has been partially supported by the NCCR Robotics through the Symbiotic Drone project, by the ML-Edge Swiss National Science Foundation (NSF) Research project (GA No. 200020182009/1), and by the ONR-G through the Award Grant No. N62909-17-1-2006.
}
\thanks{$^{1}$ U. Pale, N. Müller, A. Arza, and D. Atienza are with the Embedded Systems Laboratory of Swiss Federal Institute of Technology Lausanne, Switzerland.
         {\tt\small \{una.pale, nathan.muller, adriana.arza, david.atienza\} @ epfl.ch}}
}

\author{ Una Pale, Nathan Müller, Adriana Arza, and David Atienza $^{1}$ }

\maketitle{}

\begin{abstract}

This work presents \systemname{}, a real-time, low-complexity beat-to-beat impedance cardiography (ICG) delineation algorithm that allows hemodynamic parameters monitoring.
The proposed procedure relies only on the ICG signal compared to most algorithms found in the literature that rely on synchronous electrocardiogram signal (ECG) recordings.
\systemname{} was designed with implementation on an ultra-low-power microcontroller (MCU) in mind.
The detection accuracy of the developed algorithm is tested against points manually labeled by cardiologists. It achieves a detection Gmean accuracy of 94.9\%, 98.6\%, 90.3\%, and 84.3\% for the B, C, X, and O points, respectively. Furthermore, several hemodynamic parameters were calculated based on annotated characteristic points and compared with values generated from the cardiologists' annotations. \systemname{} achieved mean error rates of 0.11 ms, 9.72 ms, 8.32 ms, and 3.97\% for HR, LVET, IVRT, and relative C-point amplitude, respectively.

\end{abstract}


\section{Introduction}
\bstctlcite{IEEEexample:BSTcontrol}
According to the World Health Organisation (WHO), cardiovascular diseases (CVDs) are globally the highest cause of death. 17.9 million people died from CVDs in 2016, or 31\% of all global deaths~\cite{CVDsStats}. Hemodynamic parameters, such as stroke volume or cardiac output, are vital to estimate cardio-respiratory activity and evaluate the subject's condition. Hence, an ambulatory and unobtrusive monitoring of such parameters enables remote CVD monitoring and diagnosis, reduces related hospitalisation costs, expands patients’ mobility, and improves their quality of life~\cite{singhal_role_2020}.

Widely used noninvasive techniques to obtain the hemodynamic parameters are doppler echocardiography, CO2 breath analysis, seismocardiography, impedance cardiography (ICG) and phonocardiography~\cite{Dehkordi2019ComparisonInvestigation}. Among them, ICG is a suitable measurement for continuous and real-time monitoring of the hemodynamic function of the cardiovascular system since it is a noninvasive, simple, and low-cost technique. 
Many useful hemodynamic parameters can be determined from the ICG signal, such as cardiac output (CO), stroke volume (SV), overall vascular resistance, systolic time intervals~\cite{Nabian2018AnData} (e.g., left ventricular ejection time (LVET), pre-ejection period (PEP), systolic time ratio~\cite{Hafid2018FullSolutions}), or thoracic fluid content.

ICG has been used in various applications, including monitoring the cardiac rehabilitation process, monitoring of hemodynamics during hemodialysis, and pharmacological, physiological, and sleep studies~\cite{Yazdanian2016DesignMonitoring,Forouzanfar2019AutomaticCardiogram}. Moreover, with current advances in wearable technologies and ultra-low-power microcontrollers (MCU), ICG is a promising technique for monitoring hemodynamic parameters and thus, several CVDs~\cite{Mansouri2018ImpedanceDevelopments}. Still it requires an accurate and real-time detection of the ICG's characteristic points. 

However, to the best of our knowledge, no real-time, low-complexity, and beat-to-beat algorithm that can be implemented on an ultra-low-power MCU has been developed yet. 
Most of the state-of-the-art algorithms are based on the averaging of multiple cardiac cycles technique~\cite{AliSheikh2020AnAnalysis,Moissl2003FilteringCardiography,Nabian2018AnData, Sopic16DATE, Forouzanfar2018TowardCardiogram, Salah2017DenoisingPoints}, which needs the beat-to-beat synchronous reference from the electrocardiogram (ECG) signal. The assemble averaging method allows the removal of movement artifacts, respiratory influences, and stochastic distributed noise, which are the main sources of noise affecting the ICG signal. Although other ICG filtering and delineation approaches have been proposed, they are not suitable to be used in real-time and on a constrained system employing an ultra-low-power MCU.

Therefore, we propose a real-time and low-complexity procedure for the automatic detection of beat-to-beat ICG characteristic points that can be later used on ultra-low-power MCUs. Most of the automatic algorithms found in literature rely on the synchronously measured ECG signal to detect the prominent peaks of the ICG signal. In contrast, our algorithm relies only on the ICG signal, which has the advantage of reducing the complexity of acquisition and signal processing since no alignment between the two signals is needed. Moreover, our algorithm includes an adaptive filtering stage based on the state-of-the-art Savitzky-Golay filter (SG). 
 
The contributions of this work are therefore as follows:
\begin{itemize}
    \item \systemname{}: a new real-time and low-complexity beat-to-beat delineation methodology, relying only on the ICG signal for hemodynamic parameters monitoring on wearable devices. 
    
    \item An adaptive algorithm for choosing the length of the Savitzky-Golay filter, making our detection more robust to artifacts and baseline shifts.
    
    \item Adaptation of an ECG R-peak detection algorithm based on the relative energy method to detect the C points of the ICG, achieving a detection Gmean of 98.6\%. 
    
    \item A novel method to detect the pairs of X and O points, obtaining a detection Gmean of 90.3\% and 84.3\%, respectively. 

    \item Hemodynamic parameters monitoring, with precision on the level of state-of-the-art algorithms, namely, mean error rates of 0.11 ms, 9.72 ms, 8.32 ms and 3.97\% for HR, LVET, IVRT, and relative C-point amplitude, respectively.
    
    \item A new open-access database of ICG signals that include 1920 beats fully annotated by a cardiologist used to test our \systemname{} algorithm.
\end{itemize}

    
    
    

    

\section{Background and Related Work}

\subsection{Impedance Cardiography} \label{Sec:ICGdefinitions}
Impedance cardiography is a noninvasive technique for measuring changes in the thorax impedance (Z0) driven by the intrathoracic fluid changes with each heartbeat~\cite{Patterson1989FundamentalsCardiography,SherwoodChair1990MethodologicalCardiography}. Transthoracic impedance is modulated by the cardio-respiratory activity, so that a decrease in impedance is related to an increase in blood flow. ICG is measured by applying a low-amplitude, high-frequency current through two outer electrodes and then acquiring the electrical voltage difference via two inner electrodes. ICG measures ${dZ}/{dt}$, or the first derivative of the impedance signal $Z0$. 

\subsubsection{Characteristic Points Definition}
\begin{figure}[]
    \vspace{2mm}
    \centering
    \includegraphics[width=\linewidth]{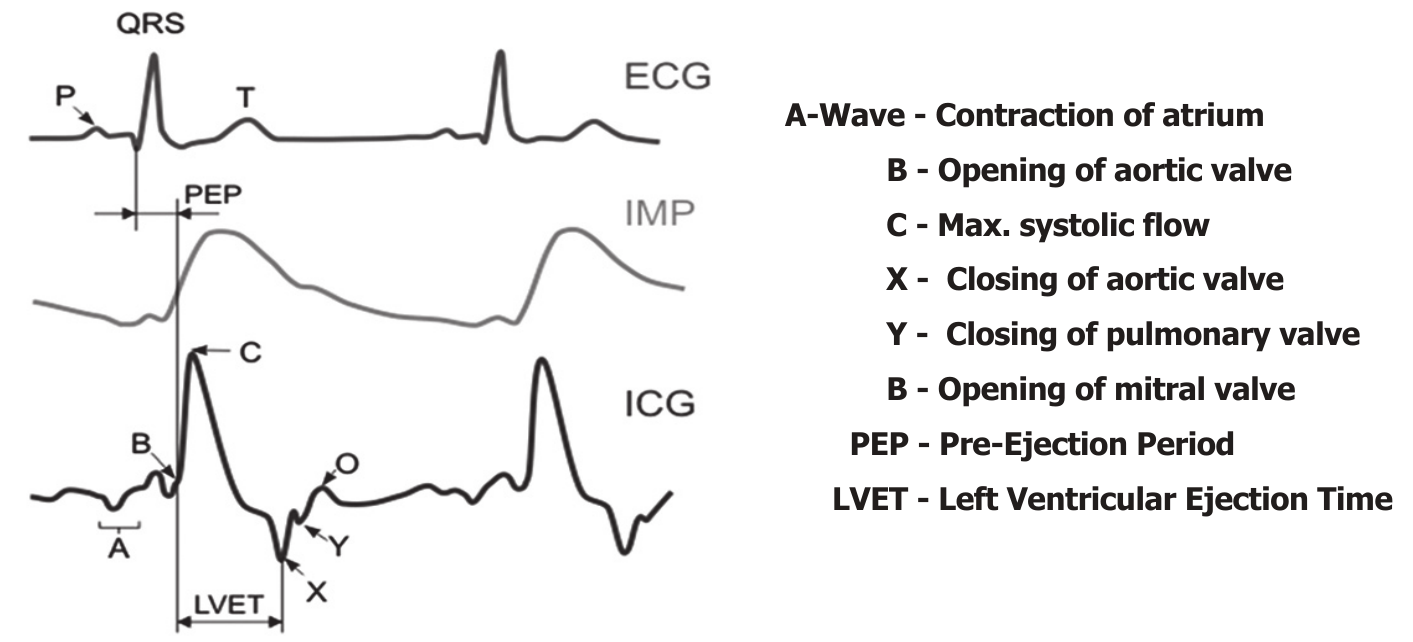}
    \caption{\small{Left: Electrocardiography (ECG), impedance variation (IMP), and impedance cardiography signal (ICG). Right: Common definition for the ICG characteristic points and hemodynamic parameters \cite{Yazdanian2016DesignMonitoring}}} 
    \label{fig:charac_points_def}
\end{figure}

The derivative of impedance ${dZ}/{dt}$ contains characteristic points related to cardio-dynamic events i.e., A, B, C, X, Y, O~\cite{Yazdanian2016DesignMonitoring, David2019}. The characteristic points of an ICG signal and its relation with cardio-dynamic events are presented in Fig.~\ref{fig:charac_points_def}. 

    \textit{C peak} -- Defined as the peak with the greatest amplitude in one cardiac cycle that represents the maximum systolic flow. 
    
    \textit{B point} -- Indicates the onset of the final rapid upstroke toward the C point~\cite{SherwoodChair1990MethodologicalCardiography} that is expressed as the point of significant change in the slope of ICG signal preceding the C point. 
    It is related to the aortic valve opening and is used to calculate the SV and the CO. However, its identification can be difficult due to variations in the ICG signal's morphology. Thus, different definitions have been proposed~\cite{Benouar2018SystematicCharacterization, DeCarvalho2011RobustC,Nabian2018AnData,Naidu2011AutomaticCardiogram}. 
    
    \textit{X point} -- Often deﬁned as the minimum ${dZ}/{dt}$ value in one cardiac cycle. However, this does not always hold true~\cite{Nabian2018AnData} due to variations in the ${dZ}/{dt}$ waveform morphology. Thus, the X point is defined as the onset of the steep rise in ICG towards the O point. It represents the aortic valve closing which occurs after the T wave end of the ECG signal.
    
   \textit{O point} -- The highest local maxima in the first half of the C-C interval. It represents the mitral valve opening.

\begin{figure}
    \vspace{2mm}
    \centering
    \includegraphics[width=\linewidth]{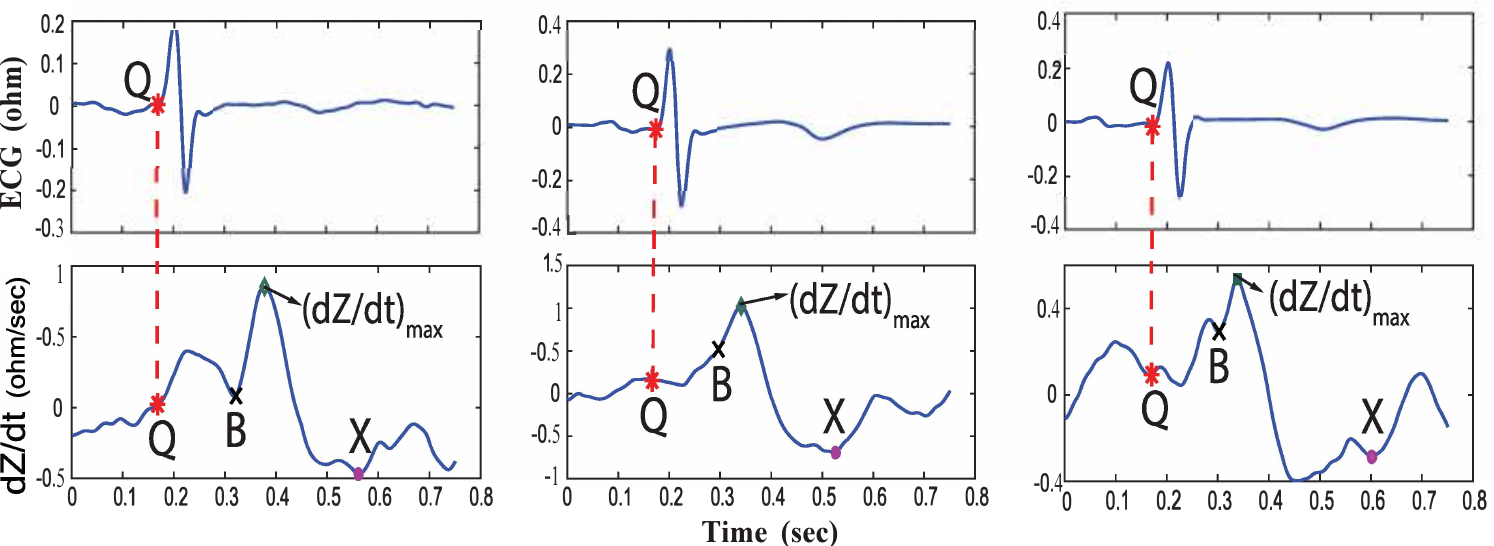}
    \caption{\small{Three examples of $dZ/dt$ (bottom) with the corresponding ECG signal (top) representing the morphological variations in the $dZ/dt$ signal with its characteristic points presented in~\cite{Nabian2018AnData}. Left case: B and X points occur at a local minimum and global minimum, respectively; Middle case: the B point is a notch; Right case: the X point is only a local minimum, not a global minimum. }} 
    \label{fig:res_morph}
\end{figure}

It is important to note that there are many variations in the morphology of the ICG signals~\cite{Benouar2018SystematicCharacterization, Nabian2018AnData}. In Fig.~\ref{fig:res_morph}, various locations of B, C and X points on different variations of ICG signal morphology can be observed. Thus, defining algorithms and detecting points on ICG signals is quite challenging.

\vspace{3pt}

\subsubsection{Hemodynamic Parameters} \label{sec:hemodyn_param}
Various hemodynamic parameters can be estimated from the ICG's characteristic points. The heart rate (HR) can be measured as the time between two consequent C peaks. The Left Ventricular Ejection Time (LVET) is the time interval measured from the B point to the X point, and can be used to estimate the stroke volume (SV) and the cardiac output (CO). The stroke volume echoes the volume of blood pumped by the left ventricle at each contraction, while the cardiac output is the total volume of blood pumped by the ventricle, usually per minute~\cite{Ipate2012THECARDIOGRAPHY}. The isovolumetric relaxation time (IVRT) is the time interval measured from the X point to the O point. The amplitude of the C peak (denoted ${dZ}/{dt}_{max}$) is commonly defined with respect to the amplitude of the corresponding B point, and is as well necessary to calculate stroke volume and cardiac output.

\subsection{State-of-the-art ICG Preprocessing and Filtering Methods} 
\label{Sec:SoA_Filter}

The most important artifacts affecting the ICG signal are sinusoidal, respiratory, muscle, and electrode artifacts~\cite{Zia2019}. Different preprocessing and filtering methods have been proposed in literature~\cite{Salah2017DenoisingPoints,David2019}, which vary from simple methods such as ensemble averaging of multiple cardiac cycles~\cite{Shyu2004TheTransform,SherwoodChair1990MethodologicalCardiography} to more advanced ones such as adaptive noise cancellation~\cite{Zia2019, Mirza2017EfficientSystems} and wavelet analysis~\cite{Stepanov2018WaveletRheocardiography,Liu2017StudyExtraction}. Among them, ensemble averaging is the most commonly used method since it eliminates stochastically distributed noise as well as respiratory influences and movement artifacts~\cite{AliSheikh2020AnAnalysis,Moissl2003FilteringCardiography,Nabian2018AnData}. 
However, averaging several cardiac cycles tends to blur less distinctive events, such as the B point, making its detection more difficult. Besides, it precludes beat-to-beat analysis of cardiac hemodynamics parameters.

\begin{figure*}[htp]
    \vspace{2mm}
    \centering
\includegraphics[width=0.90\linewidth]{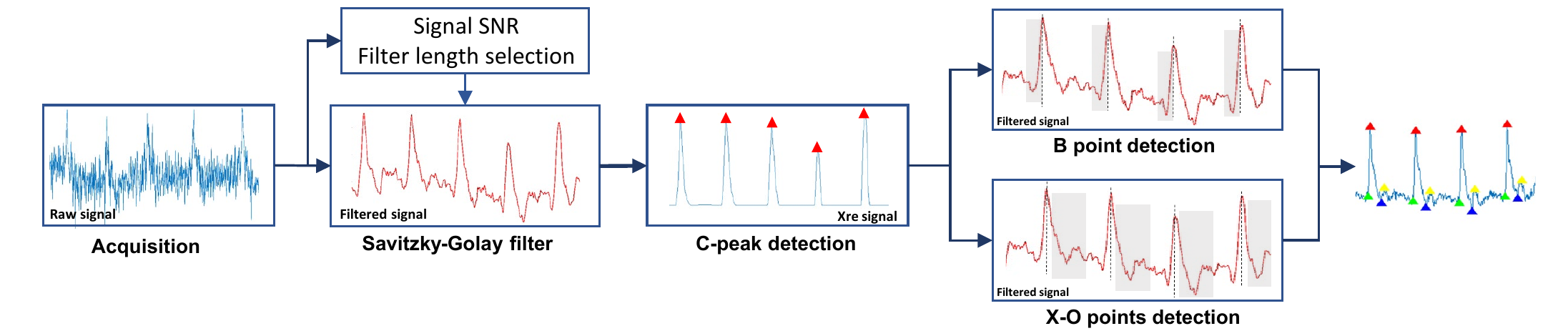}
    \caption{\small{Workflow of our proposed \systemname{} algorithm}} 
    \label{fig:workflow}
    \vspace{-3mm}
\end{figure*}

More complex techniques but also less suitable for real-time implementation on wearable devices have been proposed. For instance, adaptive filtering techniques that update the filter weights according to the statistical nature of the error signal, such as Least Mean Square and Recursive Least Square algorithms~\cite{Dromer2009ImpedanceAlgorithm}, or hybrid algorithms that have been developed to overcome the weight drift and instability of adaptive filtering algorithms~\cite{Zia2019} were tested. 

In~\cite{Salah2017DenoisingPoints}, common filtering techniques are compared, i.e. the Savitzky-Golay (SG) filter, median filter, wavelet, band-pass filter, and moving average filter. When comparing with a high-pass filter, ensemble average, and spine interpolation, the SG filter is the best option to remove respiration artifacts while preserving the beat-to-beat information~\cite{Moissl2003FilteringCardiography}.
Similarly,~\cite{Luo2005} shows that the SG filter is the most relevant for denoising ICG signals with better signal shape preservation, particularly the C peak amplitudes. Since the SG filter usage is very simple and efficient~\cite{Salah2020AutomaticProcessing}, we selected it for our methodology.

\subsection{State-of-the-art ICG Delineation Methods}

After a filtering preprocessing step, the signal is segmented, usually into beat-to-beat intervals to extract the characteristic points.
Most of the available methods, especially those using an assembling averaging step, utilize the R-peak from the ECG signal to define the beat-to-beat time interval in which to search for C peak, defined as the highest point of ICG signal.
Many of them apply known R-peak detection algorithms (e.g., the Pan-Tompkins algorithm~\cite{Salah2020AutomaticProcessing, Bagal2018}) to detect the C point due to the similarity between ECG and ICG signals.

For detection of B, X, and O points, and to address its morphological variations (see Fig.~\ref{fig:res_morph}),
several conditions based on zero crossings, minima, and maxima of the $dZ/dt$ signal and its derivatives have been proposed~\cite{Nabian2018AnData, Bagal2018, Arbol2017, Forouzanfar2019AutomaticCardiogram, Sopic16DATE, SherwoodChair1990MethodologicalCardiography}. Other algorithms employ time-frequency~\cite{Wang1995AnCardiography}, wavelet~\cite{Shyu2004TheTransform}, or machine learning-based and data-driven polynomial approaches~\cite{Forouzanfar2018TowardCardiogram}. However, they are computationally expensive with respect to the condition-based approaches. In addition, particularly time-frequency and wavelet approaches are sensitive to any artifacts that have their frequency content overlapping with the $dZ/dt$ frequency content around the characteristic points. 

As the precise identification of the B point is paramount to ensuring accurate computation of SV and CO~\cite{Mansouri2018ImpedanceDevelopments}, numerous acquisition methods have been proposed in the literature~\cite{Bagal2018}. The least computationally complex methods rely on derivatives analysis, but they are sensitive to noise and artifacts and may fail when there are several reversals, inflections, and rapid slope changes on the signal~\cite{SherwoodChair1990MethodologicalCardiography, Arbol2017}. To overcome this, in~\cite{Forouzanfar2018TowardCardiogram} and~\cite{Bagal2018}, a method is presented based on the derivatives, but, applying several conditions and corrections according to the morphology of the signal. For instance, in~\cite{Bagal2018}, the B point is detected as the sign change within 20\% and 65\% of the R–C interval window of the ICG first derivative. 
On the other hand, in~\cite{Forouzanfar2018TowardCardiogram}, the B point is found on the most prominent monotonically increasing segments between the A and C points.

X and O points detection methods are less prominent in the literature. They range from complex ones~\cite{Shyu2004TheTransform,Wang1995AnCardiography} (e.g., time-frequency analysis and wavelet-based) to simple condition checking. The simplest approach to detect the X point is as the first minimum after C wave ~\cite{Yazdani2016a, Salah2017DenoisingPoints,Salah2020AutomaticProcessing} or the lowest minimum within a given interval (i.e., in the first one-third of the C-C interval)~\cite{Forouzanfar2019AutomaticCardiogram}. However, since the X point is defined as the onset of steep rise towards the O point, some algorithms apply different conditions to detect the X-O pair~\cite{Bagal2018, Sopic16DATE}. 

Although several ICG delineation algorithms have been proposed, there is no complete proposal for low-complex and accurate delineation of B, C, X and O points that can be later implemented in a real-time wearable system. Furthermore, there are no standard evaluation metrics nor databases that allows the assessment and comparison with previously proposed methods.   

\section{Proposed \systemname{} Algorithm}

Our \systemname{} ICG delineation algorithm was developed for real-time, beat-to-beat hemodynamics parameters monitoring, which detects B, C, X and O points only relying on the ICG signal. We took advantage of many of the ideas from existing ICG delineation algorithms to improve our algorithm performance while selecting the least complex methods for a later implementation on a lightweight wearable device. The general workflow of \systemname{} is presented in Fig.~\ref{fig:workflow}. 

First, during the acquisition, the signal is segmented into windows of 3 seconds. The segments are filtered by applying the SG filter with a length adaptively selected based on the Signal to Noise Ratio (SNRs). Next, C points are detected using a novel procedure based on the R-peak REWARD algorithm~\cite{Orlandic2019REWARD:Algorithm} and designed to be implemented on ultra-low-power MCUs. Then, within the C-C intervals the B point is detected, followed by the delineation of the X-O pair, employing in both steps a derivative analysis together with condition checking following the signal morphology. 

The selected methods provide reliable detection while ensuring the low complexity of our method.
Finally, in order to have a real-time, continuous beat-to-beat analysis, the last O point found in the 3s-segmented window processed is used as the beginning of the next segmentation window.

\subsection{ICG Preprocessing using the SG filter}  \label{sec:impl_filt}

To filter the ICG signal, we use the SG filter, as it is a simple and low complexity filter that uses a polynomial approach by applying a fitting with the least squares method. However, to properly denoise the signal, the polynomial degree should be selected. Authors in~\cite{Salah2017DenoisingPoints, Salah2020AutomaticProcessing} found that the SG filter with a polynomial of order 3 has the lowest error rate for different ICG Signal to Noise Ratios (SNRs).
Further, the filter length ("the smoothing window") strongly influences filter effectiveness. Hence, selecting the appropriate filter length is essential for a correct trade-off between reducing noise and retaining important signal details needed to successfully annotate B, C, X and O points.

Adaptive SG filtering has proved its robustness and signal preservation in other applications~\cite{Li2015Real-timeAlgorithm, Acharya2016ApplicationProcessing}, and could be very useful for ICG's high SNR variability due to inter-subject variability, electrode placement, and various artifact sources.
Therefore, we propose to adaptively select the filter length by analysing its impact on the SNR of each 3s-signal window. 

Starting from a filter length of 3, we increase the length in steps of two until signal SNR reaches a target $SNR_{thr}$, or if the SNR does not improve by more than threshold $SNR_{imprThr}$. By observing how the SNR saturates above a certain filter length, $SNR_{thr}$ and $SNR_{imprThr}$ values of 30 and 1\% were chosen. Moreover, these values present a good compromise between reducing noise and maintaining a low filter length, thus reducing complexity, as well as avoiding over-smoothing of the signal (and hence potentially losing valuable details). The SNR is calculated as a ratio between the 2-norm of the high and low signal frequencies considering a 20Hz as cut-off frequency. 

\subsection{Detection of the C peaks}
The detection of the C peaks is the most important step in the detection procedure, since the other points will be located in time windows before and after the C peaks.
We have developed a new C-point detection procedure based on the "relative energy" (Rel-En) preprocessing method~\cite{yazdani_novel_2018} and an adaptation of the REWARD algorithm~\cite{Orlandic2019REWARD:Algorithm} designed to find the R peaks in the ECG.

First, the Rel-En preprocessing method is applied to the filtered ${dZ}/{dt}$ signal to enhance the C peaks. The Rel-En method considers the energies of a long sliding window (0.95s) and a short sliding window (0.14s); both centered at sample $n$. The ratio between the energies of these windows, the coefficient $c(n)$, is multiplied by $signal(n)$ resulting in a signal $X_{re}$, in which the peaks are amplified. 

The C peaks are then searched in the $X_{re}$ signal using two adaptive thresholds defined as a percentage of $Max_{val}$ in the window. $Max_{val}$ is defined as the peak amplitude value of the second biggest peak if it is significantly bigger than the mean value of the signal, otherwise the peak value of biggest peak; hence, providing robustness to the noise and outliers. 

\newfloat{algorithm}{t}{lop}
\begin{algorithm}
    \small
	\caption{B Point Detection}
	\label{alg:bpoint}
	\textbf{Input: Signal Window defined by C points} \\
	\textbf{Output: B point}
	\begin{algorithmic}[1]
		\State \textbf{def} find\_B\_point($sig, C_{pos}, C_{Ampl},A_{frac},B_{slope1},B_{slope2}$):
		\Indent
    		\State $B_{limL} = C_{pos}$ - $80ms*Fs$
    		\State $B_{limR}=$ \textbf{find} {( $sig[i] <= A_{frac} C_{Ampl}$)}
			\State $B_{min} = min(sig)$
    		\For{$i=B_{limR};i \neq B_{limL};i--$}
        		\If{$sig[i]==localMin()$ or $|sig'[i]| > B_{slope1}$}
        		    \State return $i$
        		\EndIf
    		\EndFor
    		\For{$i=B_{limR};i \neq B_{limL};i--$}
        		\If{ $sig'[i] > B_{slope2}$}
        		    \State return $i$
        		\EndIf
    		\EndFor
    		\State return $B_{min}$
		\EndIndent
		
	\end{algorithmic}
\end{algorithm}

Next, segments in which the $X_{re}$ signal goes above the upper threshold, $Thr_{max}$\textit{=0.2}$Max_{val}$, and subsequently below the lower threshold, $Thr_{min}$\textit{=0.02}$Max_{val}$, are considered active peak regions, similar to those in the REWARD approach~\cite{Orlandic2019REWARD:Algorithm}. The maximum of the signal between these two points is considered a C peak candidate. Once the peak candidates have been determined, if the peak to peak interval between two peaks is less than 0.25 seconds, only the highest of the two is kept. Further, the peaks are valid if the C-C time interval is not smaller than $CC_{min}$\textit{=1.7} times the average of the five previous C-C intervals. This condition is necessary to prevent the algorithm from detecting an O point as a C peak when, occasionally, the O point has an amplitude comparable to the one of the C peak. 

\subsection{Detection of the B points}

B point identification can be difficult due to variations in the morphology of the ICG signals (see Fig.~\ref{fig:res_morph}). Algorithm~\ref{alg:bpoint} describes our method of acquiring the B point. First, we define the time window where a B point can possibly be [$B_{limL}$,$B_{limR}$] (lines 2-3). $B_{limL}$ is defined as $C_{pos} -$80ms~\cite{Nabian2018AnData}. $B_{limR}$ is defined as the closest point before a C point where the signal amplitude is less than a fraction of C amplitude ($A_{frac}$\textit{=0.5}). 
Next, we search for either the local minimum closest to $C_{pos}$ or the first point at which the slope of the signal varies by greater than a threshold $B_{slope1}$\textit{=0.11} (lines 4-7). If $B_{limL}$ is reached before finding a B point, the search is repeated with a less strict slope threshold $B_{slope2}$\textit{=0.08} (lines 8-10). If a suitable B point is still not found, minimum $B_{min}$ is used for the B-point.

Parameters values of $A_{frac}$, $B_{slope1}$ and   $B_{slope1}$  were obtained by performing grid search for optimal parameters. For this, we randomly selected 8 blocks of 10 beats from each subject of the database and annotated by ourselves following the definition on \ref{Sec:ICGdefinitions}.

\subsection{Detection of the X and O points}

The X point is defined as the onset of the steep rise in the ICG towards the O point, which is the highest local maxima in the first half of the C-C interval. Based on this definition, we follow a new procedure on which we look for possible pairs of X-O points and select the best candidate. 
Candidates for O points are local maxima in a time window $[C_{pos}$+$CO_{min}$, $C_{pos}$+$CO_{max}]$. Similarly, candidates for X points are local minima in a window $[C_{pos} + CX_{min}$,$C_{pos} + CX_{max}]$.

Next, we look for the combinations of X-O pair candidates that meet the following criteria:

 -- O amplitude is higher than X amplitude, 
 
 -- the X-O time is within [2,15] ms range,
 
 -- there are less than 3 local minimums between O and X.

Finally, the O and X pair with the highest amplitude difference is selected as O and X points. Parameters values were obtained also by a grid search as in B point detection. We used $CO_{min}$=20 ms, $CO_{min}$=40 ms, $CX_{min}$=15 ms, $CX_{max}$=30 ms values.

\section{Experimental Setup }\label{Sec: Experimental_setup}
To assess our algorithm, we use a database manually annotated by cardiologists for the purpose of evaluating the ICG annotation algorithms. The database is published as an open-access database~\cite{ReBeatICGDatabase}. It is described below, together with the used performance metrics.

\subsection{Experimental Database} \label{Subsec:data_base}
The database includes 48 recordings of ICG and ECG signals from 24 healthy subjects during an experimental session of a virtual search and rescue mission with drones, described in~\cite{DellAgnola2020}. Two segments of 5-minute signals are selected from the first day of the reported experiment protocol from each subject. The segments correspond to a baseline state (task BL) and a higher level of cognitive workload (task CW). 
In total, the recorded database consists of 240 minutes of ICG signals.

In order to assess the performance, a subset of the database was annotated by cardiologists from the Lausanne University Hospital, which served as a test set for our algorithms. The annotation was done using an open access physiological signal labeler software \cite{labeller}. The subset consists of 4 blocks of randomly chosen signal segments containing 10 beats from BL and CW tasks of each subject. In total, 1920 (80x24) beats were annotated, each containing annotated B, C, X and O positions. 

\subsection{Performance Assessment}

To assess the performance of our proposed algorithm, the true positives (TPs), false positives (FPs), and false negatives (FNs) detected points by our algorithm were computed against the aforementioned annotations from the cardiologists. We used a tolerance of $\pm$30 ms from the annotated point~\cite{Bagal2018}. Accordingly, we use the performance metrics of sensitivity (SE), positive prediction value (PPV), detection error rate (DER), geometric mean (Gmean), mean error ($me$), and its standard deviation ($\sigma$).

Moreover, we calculated several hemodynamic parameters, both from annotation points by the cardiologists as well as from our proposed algorithm. Four hemodynamic parameters were calculated: 
\begin{enumerate}
    \item $CCtime$ interval: Related to the heart rate (HR), calculated as the interval between subsequent C points;
    \item $LVET$: Left ventricular ejection time, calculated as the time between the B and X point of the same beat; 
    \item $IVRT$: Isovolumetric relaxation time, calculated as the time between the X and O point of the same beat; 
    \item $BCampl$: Relative amplitude of C peak, which is essential for calculating stroke volume and cardiac output.
\end{enumerate}

The precision of hemodynamic parameters was measured by first matching the C peaks between automatic C peak detection and the cardiologists' annotation. Then we calculate the parameters for each of them (e.g. $LVET_{card}$ and  $LVET_{alg}$). Next, the absolute error values for all matched beats were calculated and averaged over beats and subjects and parameters. For $BCampl$, the relative error was calculated (abs($BCampl_{card}$-$BCampl_{alg}$)~/$BCampl_{card}$).

\begin{figure}[]
    \vspace{2mm}
    \centering
\includegraphics[width=0.8\linewidth]{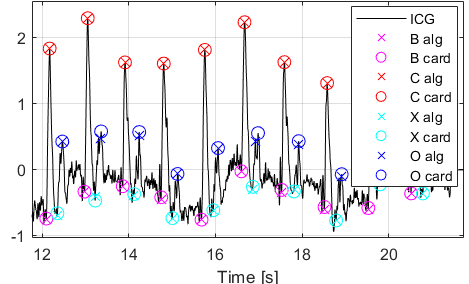}
    \caption{\small{Example of one block of data including \systemname{} automatic (X) and manual cardiologists' (O) annotation}} 
    \label{fig:annotationExample}
\end{figure}

\begin{table*}[htbp]
\vspace{2mm}
\caption{Comparison of \systemname{} with state-of-the-art methods}
\begin{center}
\begin{tabular}{|c|c|c|c|c|c|c|c|c|c|c|c|c|}
\hline
\textbf{Ref.} & \multicolumn{4}{|c|}{\textbf{B Performance measures}} & \multicolumn{4}{|c|}{\textbf{C Performance measures}} & \multicolumn{4}{|c|}{\textbf{X Performance measures}}  \\
 \cline{2-13} 
  
  & {SE [\%]} & {PPV [\%]} & {DER [\%]}  & {{$me \pm \sigma$ [ms]}}  & {SE [\%]} & {PPV [\%]} & {DER [\%]}  & {$me \pm \sigma$ [ms]} & {SE [\%]} & {PPV [\%]} & {DER [\%]}  & {$me \pm \sigma$ [ms]} \\
 
 \hline
{\cite{Bagal2018}} & --  & --  & 1.7  & 2 $\pm$ 10.5  & -- & --  & --  & 	5 $\pm$ 10  &--  &  -- & --  & 4 $\pm$ 44* \\ 
{\cite{Shyu2004TheTransform}} & -- &  -- &  -- & 3.98 $\pm$ 2.85  &--  &  -- &  -- & -- & -- &  -- & --  & 23.7 $\pm$ 14.9 \\ %
{\cite{Naidu2011AutomaticCardiogram}} & 94.4 &  93.9, &  11.7 & --  & 99.4 &  98.7    &  1.8  & --   &   97.0  &   96.5  &  6.5 & --\\ 
{\cite{Salah2017DenoisingPoints}} &  --  &  93.1  &  6.9  & --  &  --  &  100.0 &  0.0  & -- &  -- &  99.5 &  0.5 & -- \\
This work & 99.04  & 98.13 & 2.92& 2.03 $\pm$ 2.23 & 99.25  & 98.28  & 2.56  & 0.17  $\pm$ 0.73 & 98.99 & 98.43  & 2.66  & 3.29 $\pm$ 4.45 \\ 

\hline

\multicolumn{13}{l}{SE-Sensitivity. PPV- Positive  Prediction Value, DER- Detection error rate. ME-mean error and SD- standard deviation of error relative to the reference  }\\
\multicolumn{13}{l}{considering $\pm$150 ms of tolerance, as in \cite{Salah2017DenoisingPoints} and \cite{Naidu2011AutomaticCardiogram},  from  the  annotated point. * Respect to Doppler echocardiogram refences because of significant} \\ 
\multicolumn{13}{l}{ differences across the X point definitions in the literature. } \\
\end{tabular}
\label{tab1:performance_SoA}
\end{center}
\vspace{-2mm}
\end{table*}

\section{Experimental Results}

The detection performance results of our automatic \systemname{} delineation for each B, C, X and O point individually, are shown in Table~\ref{tab1:performance}. These results represent the performance of \systemname{} using as reference the beats annotated by cardiologist with $\pm$ 30 ms resolution for the matching window. Furthermore, a visual example of \systemname{} delineation and the reference annotation on one segment of signal is shown in Fig.~\ref{fig:annotationExample}. 
\systemname{} C point delineation is highly accurate, while other points delineation are quite precise as well. B point delineation is challenging due to the fact that often the correct location is defined by a slope change and not a minimum, and slope change is very sensitive to noise and filtering. 
Similarly, X and O points delineation are challenging due to many local minima and maxima sometimes present in the signal. From this reason, several conditions were necessary for identifying the correct X-O pair. Nevertheless, the \systemname{} algorithm performs very well for the B, C, X, and O points, yielding Gmean performance of 94.9\%, 98.6\%, 90.3\% and 84.3\% for each respective point. 

\begin{table}[htbp]
\caption{Performance of \systemname{} per annotated points}
\begin{center}
\begin{tabular}{|c|c|c|c|c|}
\hline
\textbf{Annot.}&\multicolumn{4}{|c|}{\textbf{Performance measures}} \\
\cline{2-5} 
\textbf{Points} & \textbf{SE [\%]}& \textbf{{PPV [\%]}} & \textbf{{Gmean [\%]}}  & \textbf{$me \pm \sigma$ [ms]} \\
\hline
B & 95.30 $\pm$ 5.65 & 94.48 $\pm$ 6.96 & 94.88 $\pm$ 6.28 & 1.75 $\pm$ 0.90  \\
C & 99.09 $\pm$ 1.86 & 98.13 $\pm$ 3.44 & 98.60 $\pm$ 2.50 & 0.12 $\pm$ 0.08  \\
X & 90.55 $\pm$ 9.51 & 90.06 $\pm$ 9.82 & 90.30 $\pm$ 9.58 & 1.09 $\pm$ 0.35  \\
O & 84.58 $\pm$ 15.45 & 84.08 $\pm$ 15.45 & 84.32 $\pm$ 15.39 & 1.31 $\pm$ 0.22  \\
\hline
\multicolumn{5}{l}{The tolerance in respect to the reference values is $\pm$ 30ms}
\end{tabular}
\label{tab1:performance}
\end{center}
\end{table}

A comparison with the existing algorithms for delineating B, C, X or O points is difficult due to different approaches used for validation. For instance, some papers compare hemodynamic parameters calculated from delineated ICG and those from echocardiogram~\cite{Bagal2018}, others in reference to ECG R peak detection~\cite{Arbol2017}, while others compare annotation positions with manual annotations~\cite{Naidu2011AutomaticCardiogram, Nabian2018AnData}. Unfortunately, almost no paper clearly mentions the resolution used for matching annotated points window, making the results hardly comparable. Further, the performance metrics are also different between papers. Nevertheless, we summarize the results from previous works in Table~\ref{tab1:performance_SoA}. Most of the papers used a resolution of $\pm$ 150ms based on \cite{AAMI2008} so we also evaluated our results with this resolution and include the results in Table~\ref{tab1:performance_SoA}. It is visible that our \systemname{} algorithm delineates all characteristic points better or on a comparable level with previous works. Among all performance measures in the table, we consider mean error $me$ [ms] to be the most reliable performance measure for comparison with different papers since it is least influenced by resolution and the one influencing the hemodynamics parameters the most. 


\begin{figure}[]
\vspace{-5mm}
    \centering
\includegraphics[width=\linewidth]{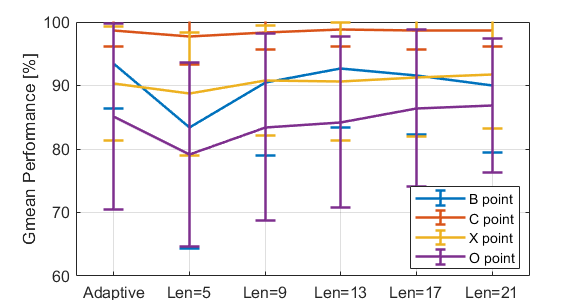}
    \caption{\small{Performance gain for our adaptive SG filter length versus fixed filter lengths for all annotated points }} 
    \label{fig:filterLebPerformance}
    
\end{figure}

Furthermore, in order to assess our adaptive filter length proposal for the SG filter providing a performance gain over a fixed length, we compare the B, C, X and O points detection performance for several fixed filter lengths versus an adaptive one as shown in Fig.~\ref{fig:filterLebPerformance}. The adaptive filter length leads to values from 5 to 25, with the mean length of 12.5 $\pm$ {5.1} over all subject data, providing better overall performance compared to equivalently sized fixed length filters.

Finally, we assessed the precision of hemodynamic parameters calculated by \systemname{} and data annotated by cardiologists. The mean error and standard deviations of absolute and relative values are expressed in Table~\ref{tab1:HemodynamicPerformance}. 
Heart rate (HR) is the most precise, with a mean error of 0.11 ms. This result is due to the fact that it depends only on C-points, whose automatic annotation was highly precise. Furthermore, LVET and IVRT are less precise, namely, having a mean error of 9.72 ms and 8.32 ms, respectively, but still well within the 30 ms resolution. 
Lastly, the relative C peak amplitude was detected with only a 3.97\% error with respect to C peak amplitude from annotated data by cardiologists.

\begin{table}[tbp]
\caption{Quality of automatic calculation of hemodynamic parameters}
\begin{center}
\begin{tabular}{|c|c|c|c|c|}
\hline
\textbf{Parameter} & \textbf{HR}& \textbf{LVET} & \textbf{IVRT}& \textbf{BCampl}\\
\hline
Mean absolute error [ms]  & 0.11$\pm$0.54  & 9.7$\pm$4.7&  8.3$\pm$9.4 &   \\
Mean relative error [\%]  & 0.01$\pm$0.04  & 3.6$\pm$1.7 &  10.2$\pm$11.0 &  3.9$\pm$6.5  \\
\hline
\end{tabular}
\label{tab1:HemodynamicPerformance}
\end{center}
\end{table}
\section{Conclusion}
In this paper, we have presented \systemname{}, a new real-time and low-complexity algorithm for beat-to-beat delineation based only on the ICG signal. In particular, the algorithm delineates 4 characteristic points, B, C, X and O whose locations are later used for calculation of hemodynamic parameters. To the best of our knowledge, this is first paper including the delineation of all the principal characteristic points of the ICG for hemodynamic parameters monitoring, that allows a later implementation on ultra-low-power MCU.  

Our algorithm employed three main state-of-the-art techniques that have both high reliability and low complexity. First, to filter the signal, we employed the Savitzky-Golay filter that is easily implemented as a convolution of the signal with the filter's coefficients, which are integers calculated offline. Second, we used the relative energy method for a highly precise C-peak detection allowing further beat-to-beat delineation. This method has been previously implemented on low-power MCU. Finally, for the delineation of B, X and O points we used the derivative analysis with conditioning checking based on time-intervals and signal-amplitude, being a low-complexity method that has also shown high performance.

The automatic detection procedure was compared against the annotated points provided by cardiologists and yielded Gmean performance of 94.9\%, 98.6\%, 90.3\%, and 84.3\% for B, C, X, and O points, respectively. 
Finally, we have assessed the precision of several  hemodynamic parameters, obtaining mean errors of 0.11 ms, 9.72 ms, 8.32 ms, and 3.97\% for HR, LVET, IVRT, and relative C-point amplitude, respectively. 
These overall results indicate that our proposed algorithm is highly precise, and could be implemented on a low-power wearable device for hemodynamic parameter monitoring. \\

\section*{ACKNOWLEDGMENT}
We especially want to thank Dr. David Meier and Prof. Olivier Muller for all the time and expertise in annotating the ICG database on which we evaluated proposed \systemname{} algorithm.


\def\url#1{}

\bibliographystyle{IEEEtran}
\bibliography{manualReferences.bib}

\end{document}